\documentclass[12pt]{iopart}

\usepackage{graphicx,color}

\begin{document}

\paper[]{Rearrangement of cluster structure during fission processes}

\author{A G Lyalin\dag\ddag, 
        O I Obolensky\S,  
        A V Solov'yov\ddag\S, 
        Il A Solov'yov\ddag\S\ 
        and W Greiner\ddag  }

\address{\dag\ Institute of Physics, St Petersburg State University, 
Ulianovskaya str. 1, 198504 St Petersburg, Petrodvorez, Russia}  

\address{\ddag\ Institut f\"{u}r Theoretische Physik der Universit\"{a}t
Frankfurt am Main, Robert-Mayer Str. 8-10, D-60054 Frankfurt am Main, Germany}

\address{\S\ A.F. Ioffe Physical-Technical Institute, Russian
Academy of Sciences, Politechnicheskaja str. 26, 194021 St Petersburg,  Russia}

\ead{lyalin@th.physik.uni-frankfurt.de}

\begin{abstract}
Results of molecular dynamics simulations of fission
reactions $Na_{10}^{2+} \to Na_7^+ + Na_3^+$ and $Na_{18}^{2+} \to 2 Na_9^+$
are presented.
Dependence of the fission barriers on isomer structure
of the parent cluster is analyzed.
It is demonstrated that the energy necessary for removing
homothetic groups of atoms from the parent cluster
is largely independent of the isomer form of the parent cluster.
Importance of rearrangement of the cluster structure during the
fission process is elucidated. This rearrangement may include
transition to another isomer state of the parent cluster
before actual separation of the daughter fragments begins and/or
forming a "neck" between the separating fragments.
\end{abstract}


\maketitle


Fission of charged atomic clusters occurs when repulsive Coulomb forces,
arising due to the excessive  charge, overcome the electronic binding energy
of the cluster \cite{Sattler81,Naher97,Yannouleas99}.
This mechanism of the cluster
fission is in a great deal similar to the nuclear fission phenomena.
Experimentally, multiply charged metal clusters
can be observed in the mass spectra
when their size exceeds the critical size of stability, which depends
on the metal species and cluster charge \cite{Brechignac89,Brechignac94,Martin84}.

We report the results of the {\it ab initio}
molecular dynamics (MD) simulations of the fission processes
$Na_{10}^{2+} \to Na_7^+ + Na_3^+$ and
$Na_{18}^{2+} \to 2 Na_{9}^+$. Both symmetric and asymmetric fission channels are considered.
We have investigated the
parent cluster isomer dependence of the fission barrier
for the reaction $Na_{10}^{2+} \to Na_7^+ + Na_3^+$.
To the best of our knowledge, a comparative study
of fission barriers for various isomers by means of
quantum chemistry methods has not been carried out before.
Note that such a study is beyond the scope of
simpler approaches
which do not account for ionic structure of a cluster.

We found that 
{ the direct separation barrier for the reaction 
$Na_{10}^{2+} \rightarrow Na_{7}^{+} + Na_{3}^{+}$}
has a weak dependence
on the isomeric structure of the parent cluster.
We note, however, that the groups of atoms to be removed
from the parent cluster isomers must be chosen with care;
one has to identify homothetic groups of atoms in each
fissioning isomer.
The weak dependence on the isomeric state of the parent 
{ $Na_{10}^{2+}$} cluster
implies that
the particular ionic structure of the cluster
is largely insignificant for the shape and
height of the fission barrier.
This is due to the fact that the 
maximum of fission barriers in considered cases 
are located at distances
comparable or exceeding
the sum of the resulting fragments radii.
At such distances the interaction between the fragments, apart from
the Coulombic repulsion,
is mainly determined by the electronic properties rather
than by the details of the ionic structure of the fragments.
This is an important argument for justification
of the jellium model approach to the description
of the fission process of multiply charged metal clusters.

We have demonstrated the importance of rearrangement of the cluster
ionic structure during the fission process.
The possibility of rearrangement of the cluster structure
leads to the fact that direct fission of a cluster isomer
in some cases may not be the energetically optimum path
for the fission reaction. Alternatively, the reaction can go through
transition to another isomer state of the parent cluster.
This transition can occur in the first phase of the fission
process, before separation of the fragments actually begins.
We show that this is the case for the fission of $C_{4v}$ and $D_{4d}$
isomers of $Na_{10}^{2+}$ cluster. 

The rearrangement of ionic structure
may be important
also after the fragments began to separate.
For $Na_{18}^{2+} \to 2 Na_9^+$ reaction,
two magic fragments $Na_9^+$ form
a metastable transitional state in which the fragments are
connected by a "neck". This "necking" allows for significant
reduction in the height of the fission barrier.
Note that the similar necking phenomenon is known for the nuclear fission process 
\cite{Eisemberg_and_Greiner}.


In our molecular dynamics simulations we utilize methods
of density functional theory (DFT).
Within the DFT one has to solve the Kohn-Sham equations
\cite{Kohn-Sham}
\begin{equation}
\left(
\frac{p^2}{2}+U_{\rm i}+V_{\rm H}+V_{\rm xc}
\right)
\psi_i = \varepsilon_i \psi_i,
\end{equation}
\noindent where the first term corresponds to the kinetic
energy of an electron 
{ from the Kohn-Sham reference system}, 
$U_{\rm i}$ describes the attraction of the $i^{\rm th}$ electron to the nuclei
in the cluster, $\psi_{\rm i}$ is the electronic orbital,
$V_{\rm H}$ is the Hartree part of the
inter-electronic interaction,
\begin{equation}
V_{\rm H}({\bf r})=\int
\frac{\rho({\bf r}^\prime)}{|{\bf r}-  {\bf r}^\prime |}
{\rm d} {\bf r}^\prime ,
\end{equation}
\noindent $\rho({\bf r})$ is the electron density,
$V_{\rm xc}$ is the local exchange-correlation potential
defined as the functional derivative
of the exchange-correlation energy functional
\begin{equation}
V_{\rm xc} = \frac{\delta E_{\rm xc} [\rho]}{\delta \rho({\bf r})},
\end{equation}
\noindent where the exchange-correlation energy is partitioned
into two parts, referred to as {\it exchange} and {\it correlation}
parts:
\begin{equation}
E_{\rm xc} [\rho]= E_{\rm x}(\rho) + E_{\rm c} (\rho).
\end{equation}
\noindent Physically, these two terms correspond to same-spin and
mixed-spin interactions, respectively. Both parts are
functionals of the electron density, which can be of two distinct
types: either a local functional depending on the
electron density $\rho$ only or a
gradient-corrected functional depending
on the electron density and its gradient $\nabla \rho$.

There is a variety of exchange-correlation functionals
in the literature.
We have used the three-parameter Becke-type gradient-corrected
exchange functional with the gradient-corrected correlation
functional of Lee, Yang, and Parr (B3LYP) \cite{Becke93}.
For the explicit form of this functional we refer to the original papers
\cite{Becke88,VWN80,LYP88}.
The B3LYP functional has proved to be a reliable tool for
studying the structure and properties of small metal clusters.
It provides high accuracy at comparatively low computational
costs. For a discussion and a comparison with other approaches,
see \cite{struct_Na,struct_Mg}.
Note that the density of the parent cluster and two daughter 
fragments (including the overlapping region before scission point) 
almost does not change during the fission process 
(by analogy with the deformed jellium model, 
see \cite{HFLDA1,HFLDA2} for more details). 
This means that the B3LYP method works adequately for any fragment separation 
distances, $d$, during the fission process.

The calculations have been carried out with the use of
the GAUSSIAN 98 software package \cite{Gaussian}.
The 6-311G(d) and LANL2DZ basis sets of primitive Gaussian functions
have been used to expand the cluster orbitals \cite{Chemistry}.
The 6-311G(d) basis has been used for simulations involving
$Na_{10}^{2+}$ cluster. This basis set takes into account electrons from all atomic
orbitals, so that the dynamics of all particles in the system
is taken into account.
For $Na_{18}^{2+}$ cluster we have used more numerically efficient
LANL2DZ basis, for which valent atomic electrons move in
an effective core potential (see details in \cite{Chemistry}).

{
To simulate the fission process 
we start from the optimized geometry of a cluster 
(for details of the geometry optimization procedure 
see \cite{struct_Na,struct_Mg})
and choose the atoms the resulting fragments would consist of. 
The atoms chosen for a smaller fragment 
are shifted from their locations in the parent cluster
to a certain distance.
Then, the multidimensional
potential energy surface, its gradient and forces with respect to the molecular coordinates
are calculated. These quantities specify the direction along the surface 
in which the energy decreases the most rapidly and provide information 
for the determination of the next step for the moving atoms.
If the fragments are removed not far enough
from each other then
the attractive forces prevailed over the repulsive ones and 
the fragments stuck together forming the unified cluster again.
In the opposite situation the repulsive forces dominate
and the fragments drift away from each other.
The dependence of the total energy of the system on the
fragment separation distance forms the fission barrier.
The aim of our simulations is to find the fission pathway 
corresponding to the minimum of the fission barrier.}



\begin{figure}[h]
\vspace*{0.5cm}
\begin{center}
\hspace*{1.5cm}\includegraphics[scale=1]{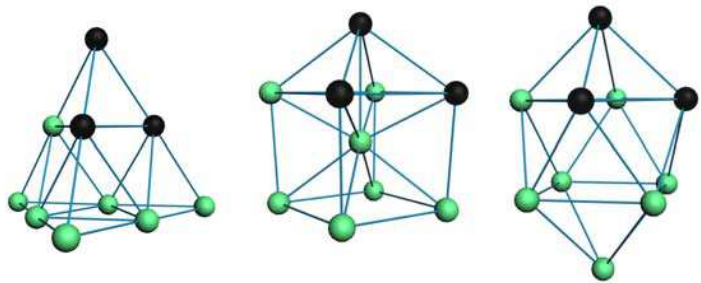}
\end{center}
\caption{Three isomers of $Na_{10}^{2+}$ cluster. From left to right:
the ground state isomer of distorted $T_d$ point symmetry group
(total energy is -1622.7063 a.u.);
an isomer of $C_{4v}$ point symmetry group (total energy is -1622.6888 a.u.,
that exceeds the lowest energy state by 0.476~eV);
an isomer of $D_{4d}$ point symmetry group (total energy is -1622.6860 a.u.,
that exceeds the lowest energy state by 0.553~eV).
{ The homothetic group of three atoms marked by black color.}}
\label{isomers}
\end{figure}

There are usually many stable isomers of a cluster
with energies slightly exceeding the energy of the ground state isomer.
In order to analyze the isomer dependence of the fission
barrier in the reaction  $Na_{10}^{2+} \to Na_{7}^{+} + Na_{3}^{+}$
we have picked two energetically low-lying isomers with
the point symmetry groups $C_{4v}$ and $D_{4d}$
differing from the distorted $T_{d}$ point symmetry group
of the ground state parent $Na_{10}^{2+}$ cluster. Three isomer states
of the $Na_{10}^{2+}$ cluster are shown in figure~\ref{isomers}.

\begin{figure}[h]
\begin{center}
\includegraphics{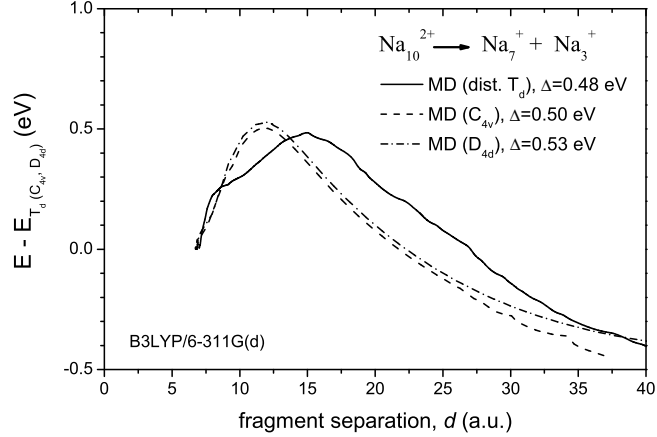}
\end{center}
\caption{Fission barriers for {  
separating the homothetic group of three atoms 
(marked by black color in figure~\ref{isomers})} from
three isomers of $Na_{10}^{2+}$ cluster
derived from molecular dynamics simulations
(direct $Na_{10}^{2+} \to Na_7^+ + Na_3^+$
fission channel).
The barriers plotted versus
distance between the centers of mass of the fragments.
Solid, dashed, and dashed-dotted lines correspond to
distorted $T_d$, $C_{4v}$, and $D_{4d}$ point symmetry groups
isomers of the parent cluster, respectively.
{Energies are measured from the energy of the 
ground state of the corresponding isomers, i.e. we plot 
$E - E_{T_d (C_{4v}, D_{4d})}$, where $E$ is the total energy of the system and 
$E_{T_d (C_{4v}, D_{4d})}$ are the ground energies of the $T_d$, $C_{4v}$ and $D_{4d}$ 
isomer states of the parent $Na_{10}^{2+}$ cluster, respectively. 
}}
\label{isobarriers}
\end{figure}

In figure~\ref{isobarriers} we show fission barriers for separation three atoms
from the $C_{4v}$, $D_{4d}$, and $T_{d}$ isomers of the $Na_{10}^{2+}$ cluster.
In this figure zero level of energy is chosen for each parent isomer
separately and corresponds to the minimum of total energy of that isomer.
The initial distances between the centers of mass of two (future)
fragments are finite so that the barriers do not start at the origin.

The barriers for all three channels are close. 
The weak sensitivity of the fission barrier
on the isomeric states of the reactants
can be explained if one notices that the barrier maxima 
are located at distances comparable to or exceeding
the sum of the resulting fragments radii, that is
not far from the scission point.
At such distances the interaction between the fragments,
apart from Coulombic repulsion,
is mainly determined by the electronic properties rather
than by the details of the ionic structure of the fragments.
This is an important argument for justification
of the jellium model approach to the description
of the fission process of multiply charged metal clusters.

It is important to note that
the barriers presented in figure~\ref{isobarriers}
are calculated in assumption that fission occurs
for the fixed (given) isomers. However,
since $C_{4v}$ and $D_{4d}$ isomers
are not the lowest energy states of $Na_{10}^{2+}$ system,
there could be other processes competing with fission.
One of such processes is rearrangement of the cluster structure.


Rearrangement of the cluster structure during the fission process
may significantly reduce the fission barrier.
Such rearrangement may occur before the actual separation
of the daughter fragments begins or after that.

\begin{figure}[h]
\begin{center}
\includegraphics{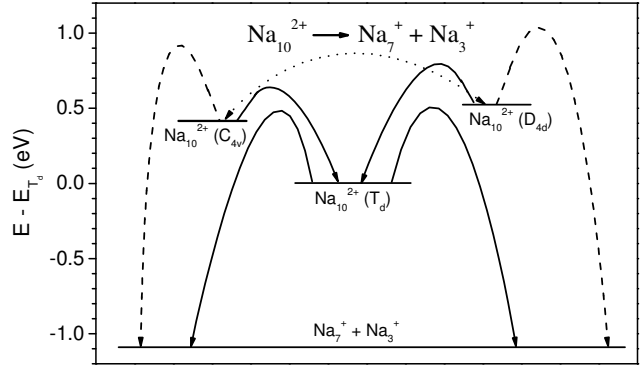}
\end{center}
\caption{Energy levels of some states of the $Na_{10}^{2+}$ system
and schematic barriers for transitions between these states.
Energies are measured from the energy of the ground $T_d$ state of
the $Na_{10}^{2+}$ cluster.}
\label{levels}
\end{figure}

\begin{figure}[h]
\begin{center}
\includegraphics[scale=1]{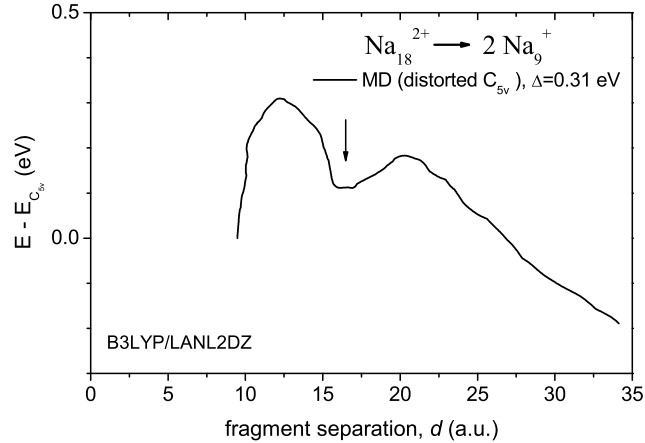}
\end{center}
\caption{Fission barrier for $Na_{18}^{2+} \to 2 Na_9^+$
channel derived from molecular dynamics simulations as a function
of distance between the centers of mass of the fragments.
{ Energy is measured from the energy of the ground $C_{5v}$ state of
the $Na_{18}^{2+}$ cluster. }
The arrow shows position of the meta-stable transitional
state, see also figure~\protect{\ref{rearrangement}}. This results is in a good agreement
with the results of the jellium model \cite{HFLDA1,HFLDA2}.}
\label{fission18}
\end{figure}
\begin{figure}[h]
\begin{center}
\hspace*{2.cm}\includegraphics[scale=0.4]{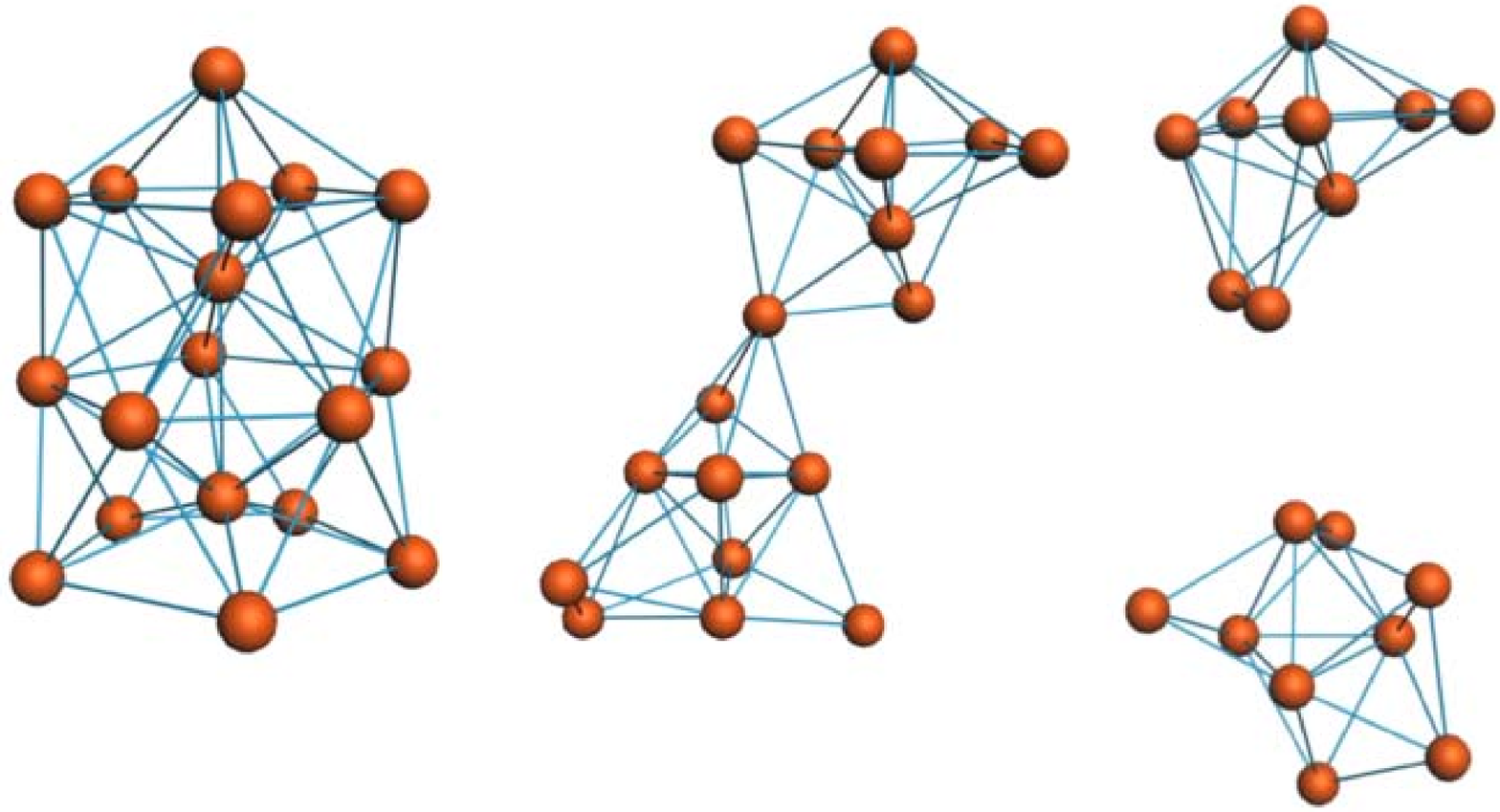}
\end{center}
\caption{Rearrangement of the cluster structure 
during the fission process $Na_{18}^{2+} \to 2 Na_9^+$.
From left to right: ground state of the parent cluster;
"necking" between the two fragments leads to a meta stable
intermediate state and significantly reduces the fission barrier height;
two $Na_9^+$ fragments drifting away from each other.}
\label{rearrangement}
\end{figure}

Fission of $C_{4v}$ and $D_{4d}$ isomers of $Na_{10}^{2+}$ cluster
is an example of situation where rearrangement of the cluster structure
takes place before the fragments start to separate.
In figure~\ref{levels} we show schematically the total energies of the  $Na_{10}^{2+}$
$T_d$, $C_{4v}$ and $D_{4d}$ isomers and barriers for the transitions between
those states. It is seen from the figure that
transition to the ground ($T_d$) state with subsequent fission
into the $Na_3^+$ and $Na_7^+$ fragments,
$Na_{10}^{2+}(C_{4v} {\rm\ or\ } D_{4d}) \to Na_{10}^{2+}(T_{d}) \to Na_7^+ + Na_3^+$,
(shown by solid lines) is preferred path for
fission of both $C_{4v}$ and $D_{4d}$ isomers of the $Na_{10}^{2+}$
cluster and requires only about 0.2~eV for $C_{4v}$ isomer
and 0.26~eV for $D_{4d}$ isomer.
In contrast, the direct fission process,
$Na_{10}^{2+}(C_{4v} \rm{\ or\ } D_{4d}) \to Na_7^+ + Na_3^+$,
(shown by dashed lines) requires about 0.5~eV.
We also show the barrier for the transition between the $C_{4v}$ and $D_{4d}$ isomers.

Another example of cluster structure rearrangement
in the fission process is the $Na_{18}^{2+} \to 2 Na_9^+$ reaction.
The fission barrier for this reaction is shown in figure~\ref{fission18}.
If two fragments of the parent cluster were not allowed
to adjust their ionic structure the fission barrier would be
about 1~eV. Rearrangement of the cluster structure
allows to reduce the fission barrier down to 0.31~eV.
During the fission process the daughter fragments
start to drift away
from each other and a "neck" forms between the fragments.
Formation of the "neck" results in a metastable transitional state.
The geometry of this state, as well as the geometry of the parent
cluster are shown in figure~\ref{rearrangement}.

In table~\ref{tab} we have summarized our results 
for the fission barrier heights and compared them with the results
of other molecular dynamics simulations and with the predictions
of the jellium model.

\begin{table}
\caption{\label{tab} Summary of the fission barrier heights (eV).}
\begin{indented}
\item[]\begin{tabular}{@{}lll}
\br
                  & $Na_{10}^{2+} \to Na_7^+ + Na_3^+$ & $Na_{18}^{2+} \to 2 Na_9^+$\\
\mr
MD (this work)    & 0.49 (distorted $T_d$)             & 0.31 \\
MD\cite{Montag95a}& 0.67                               & 0.52 \\
MD\cite{Guet01}   & 0.54                               & ---  \\
Jellium model\cite{HFLDA1,HFLDA2}   & 0.16             & 0.48 \\
\br
\end{tabular} 
\end{indented}
\end{table}


We have investigated two aspects of charged metal cluster
fission process: dependence of the fission barrier
on isomer state of the parent cluster and importance
of rearrangement of the cluster ionic structure
during the fission process.

We found that for a consistent choice of the atoms removed
from the cluster the fission barrier 
{ for the reaction $Na_{10}^{2+} \to Na_7^+ + Na_3^+$}
has a weak dependence
on the initial isomer structure of the parent cluster.
This implies that
the particular ionic structure of the cluster
is largely insignificant for the height of the fission barrier.
This is an important argument for justification
of the jellium model approach to the description
of the fission process of multiply charged metal clusters.

We have shown importance of rearrangement of the cluster
ionic structure during the fission process.
The fission reaction can go through
transition to another isomer state of the parent cluster.
This transition can occur
before actual separation of the fragments begins and/or
"neck" between the separating fragments is formed.
In any case the resulting fission barrier can be significantly
lower compared to the one for the direct fission path.

\ack

The authors acknowledge support of this work by 
the Alexander von Humboldt Foundation, 
DFG, the Studienstiftung des deutschen Volkes,
{INTAS}, 
Russian Foundation for Basic Research {(grant No 03-02-16415-a)}, 
Russian Academy of Sciences {(grant 44)} 
and the Royal Society of London.

\end{document}